\documentclass[fleqn,usenatbib]{mnras}

\usepackage{newtxtext,newtxmath}

\usepackage[T1]{fontenc}
\usepackage{ae,aecompl}
\usepackage{times}
\usepackage{graphicx,subfigure,float}

\usepackage{graphicx}	
\usepackage{amsmath}	
\usepackage{amssymb}	
\usepackage{color}

\usepackage{titlesec}
\titlespacing*{\section}{0pt}{0.95\baselineskip}{\baselineskip}

\newcommand{\tcb}{\textcolor{black}}

\title[Marked Correlation Function in MG]{Testing modified gravity using a marked correlation function}

\author[Armijo et al.]{
Joaqu\'\i n Armijo,$^{1}$\thanks{E-mail: jarmijo@astro.puc.cl}
Yan-Chuan Cai,$^{2}$
Nelson Padilla,$^{1}$
Baojiu Li$^{3}$ and
John A. Peacock$^{2}$
\\
$^{1}$Instituto de Astrof\'\i sica, Pontificia Universidad Cat\'olica de Chile, Vicu\~na Mackenna 4860, Santiago, Chile\\
$^{2}$Royal Observatory Edinburgh,  Blackford Hill, Edinburgh EH9 3HJ, United Kingdom\\
$^{3}$Institute for Computational Cosmology, Durham University, Durham, DH1 3LE, United Kingdom
}

\date{Accepted XXX. Received YYY; in original form ZZZ}

\pubyear{2017}

\begin{document}
\label{firstpage}
\pagerange{\pageref{firstpage}--\pageref{lastpage}}
\maketitle

\begin{abstract}
In theories of modified gravity with the chameleon screening mechanism, the strength of the fifth force depends on environment. This induces an environment dependence of structure formation, which \tcb{differs from} $\Lambda$CDM. We show that these differences can be captured by the marked correlation function. With the galaxy correlation functions and number densities calibrated to match between $f(R)$ and $\Lambda$CDM models in simulations, we show that the marked correlation functions from using either the local density or halo mass as the marks encode \tcb{extra} information, \tcb{which can be used to test these theories}. We discuss possible applications of these statistics in observations. 
\end{abstract}




\section{Introduction}

Theories of modified gravity were introduced as alternatives to the $\Lambda$-cold-dark-matter ($\Lambda$CDM) paradigm to explain the late-time cosmic acceleration. 
In light of the recent detection of gravitational waves from the binary neutron star merger GW170817 and simultaneous measurement of its optical counterpart GRB170817A, several popular classes of model are ruled out \citep[e.g.][]{Lombriser2016,Baker2017, Sakstein2017,Ezquiaga2017,Creminelli2017}, although many other models remain viable and would affect the growth of large-scale structure, such as Brans-Dicke type theories including $f(R)$ gravity \citep{Felice2010}, derivative-coupling theories including the normal-branch Dvali-Gabadadze-Porrati (nDGP) model \citep{Dvali2000}, and more complex variants of dark energy within standard gravity. It remains important to test the equivalence principle and General Relativity (GR) at cosmological scales.

A general feature of the surviving modified gravity models is that they often rely on  screening mechanisms to suppress the fifth force in 
high density regions. This is true for both the $f(R)$ \citep{Li:2007xn, Brax:2008hh} and nDGP models \citep{Dvali2000}. 
The former features a chameleon screening and the latter the Vainshtein screening mechanism \citep{Khoury2004,Vainshtein1972}. 
This inevitably alters structure formation in an environmental dependent manner, i.e. in the regime where the fifth force is suppressed, gravity is back to GR and structure formation remains similar to that of the $\Lambda$CDM; in the places where the fifth force is unscreened, such as in low density regions in the $f(R)$ model, or outside the Vainshtein radius in nDGP model, the additional fifth force acts to change structure formation in a complex way. This provides opportunities to test these models using statistics that are sensitive to the environment-dependent nature of structure formation. In this letter, we explore using the marked correlation method to test gravity using the $f(R)$ model as an example, motivated by the methodology proposed in \cite{White2016}.

The marked correlation is a high order statistical method which \tcb{contains information beyond the galaxy two point correlation function.} It is useful for studying the connections between properties of galaxies, such as luminosity and environmental density, with their spatial clustering with the flexibility of the choice of the mark \citep[e.g.][]{Beisbart2000,Sheth2004, Harker2006, Wechsler2006}. This statistic has been applied to break degeneracies between the halo occupation and $\sigma_8$ in two different cosmological models with the same clustering \citep{white2008}. The same principle should be \tcb{applicable} to distinguish MG and $\Lambda$CDM \citep{White2016}. In this letter, using galaxy catalogues from both $f(R)$ and $\Lambda$CDM models that are tuned to have the same clustering, we explore different mark statistics to see if these models can be told apart.

The key question is what mark is the optimal to fulfill our task. We explore two quantities, local density and halo mass, as the mark, which we believe should serve best for our purpose of capturing the difference due to the distinct environmental dependencies for structure formation in $f(R)$ and $\Lambda$CDM models.
The outline of this letter is the following: In \S~\ref{sec:theory} we describe $f(R)$ theory and our simulations. The results of the marked correlation function are shown in \S~\ref{sec:results}. We draw conclusions and discuss our results in \S~\ref{sec:conc}. 

\section{Theory and simulations}\label{sec:theory}

\subsection{The \textit{f(R)} model of gravity}

The MG model studied in this letter is $f(R)$ gravity see \cite{Felice2010} for a review, which extends GR by including a function of the Ricci scalar $R$, $f(R)$, in the Einstein-Hilbert action:
\begin{equation}
S = \int{\rm d}^4x \sqrt{-g} \left\{  \frac{1}{2\kappa^2}[R+f(R)] + \mathcal{L}_m \right\},
\end{equation}
where $\kappa
^2=8\pi G$, $G$ is Newton's constant, and $g$ is the determinant of the metric $g_{\mu\nu}$. 
In this model, gravity between massive particles is governed by a modified Poisson equation:
\begin{equation}\label{eq:Phi}
\vec{\nabla}^2 \Phi = \frac{16\pi G}{3}a^2\left[\rho_m-\bar{\rho}_m\right] + \frac{1}{6}a^2[R(f_R)-\bar{R}],
\end{equation}
in which $\rho_m=\rho_m({\bf x},t)$ is the density of non-relativistic matter at scale factor $a$, an overbar means the cosmic mean of a quantity and $f_R\equiv{\rm d}f(R)/{\rm d}R$ is an additional scalar degree of freedom (a scalar field) which is governed by an equation of motion (EoM):
\begin{equation}\label{eq:fR}
\vec{\nabla}^2 f_R = -\frac{1}{3}[R(f_R) - \bar{R} + 8\pi G(\rho_m - \bar{\rho}_m)].
\end{equation}
Eqs.~\eqref{eq:Phi} and \eqref{eq:fR} can be combined to obtain
\begin{equation}\label{eq:Phi2}
\vec{\nabla}^2 \Phi = 4\pi Ga^2\left[\rho_m-\bar{\rho}_m\right] - \frac{1}{2}\vec{\nabla}^2f_R,
\end{equation}
which indicates that $-\frac{1}{2}f_R$ can be considered as the potential of a force, called the {\it fifth force}, that is mediated by the scalar field $f_R$. 

An interesting feature of this model is the chameleon screening mechanism \citep{Khoury2004}. 
Inside a deep Newtonian potential (e.g., the solar system) or with a uniform high matter density (e.g., the early Universe), the solution to Eq.~\eqref{eq:fR} is dynamically driven to $|f_R|\rightarrow0$ so that Eq.~\eqref{eq:Phi2} reduces to the standard Poisson equation: in this regime GR is recovered, hence offering a way for the theory to pass stringent solar system tests of gravity.

In contrast, in shallow Newtonian potentials, the dynamics of Eq.~\eqref{eq:fR} is such that $\delta R = R - \bar{R}$ is negligible, and Eq.~\eqref{eq:Phi2} reduces to
\begin{equation}\label{eq:Phim}
\vec{\nabla}^2\Phi = \frac{16}{3}\pi Ga^2\left[\rho_m-\bar{\rho}_m\right],
\end{equation}
indicating a $1/3$ enhancement of gravity w.r.t.~GR, or a fifth force with $1/3$ the strength of standard gravity at maximum, independent of the form of $f(R)$. This fifth force can enhance the growth of dark matter haloes \citep{Cai2015}, and make cosmic voids grow larger by evacuating more matter from void centres \citep{Clampitt2013}.
The fact that the fifth force is strong in low-density regions but suppressed in high-density regions implies that the difference from GR can be strengthened by up-weighting low density regions using marked statistics, thus offering a way to distinguish the model from $\Lambda$CDM. We shall show this is the case next, and for illustration we
adopt the form of $f(R)$ proposed in \cite{HS2007}:
\begin{equation}
f(R) = -m^2\frac{c_1(-R/m^2)^n}{c_2(-R/m^2)^n + 1},
\end{equation}
where $m^2 = \kappa^2\bar{\rho}_0/3$,  $\bar{\rho}_0$ being the mean density of the Universe today. 

For a realistic expansion history, $|R|\gg m^2$ for $z\geq0$, so that 
\begin{equation}\label{eq:exp}
f(R) \approx -\frac{c_1}{c_2}m^2 + \frac{c_1}{c_2^2}m^2\left( \frac{m^2}{R} \right)^n,
\end{equation}
to a good approximation. 
If we set ${c_1}/{c_2} = 6{\Omega_\Lambda}/{\Omega_m}$, where $\Omega_m$ is the density parameter for matter today and $\Omega_\Lambda=1-\Omega_m$, the model can accurately mimic a $\Lambda$CDM expansion history. Meanwhile,
\begin{equation}\label{eq:sca} 
f_R \approx -n\frac{c_1}{c_2^2}\left( \frac{m^2}{R} \right)^{n+1},
\end{equation}
which can be inverted to find $R(f_R)$ which is used in Eqs.~(\ref{eq:Phi}, \ref{eq:fR}).
Thus the model has two free parameters, $n$ and $c_1/c_2^2$, 
which can be related to the value of $f_{R0}$ today by using Eq.~\eqref{eq:sca}:
\begin{equation}
\frac{c_1}{c_2^2} = -\frac{1}{n}\left[ 3 \left( 1 + 4\frac{\Omega_{\Lambda}}{\Omega_m} \right) \right]^{n+1} f_{R0}.
\end{equation}
A smaller $|f_{R0}|$ means weaker deviation from GR. The current {cosmological} constraint on these parameters is and $|f_{R0}| \lesssim 10^{-5}$ \citep[e.g.][]{Cataneo2015,Liu2016}; we fix $n=1$ in this work. 

\subsection{Simulations and mock galaxy catalogues}
\label{sec:simulation}
\begin{figure}
\centering
\includegraphics[width=.85\linewidth]{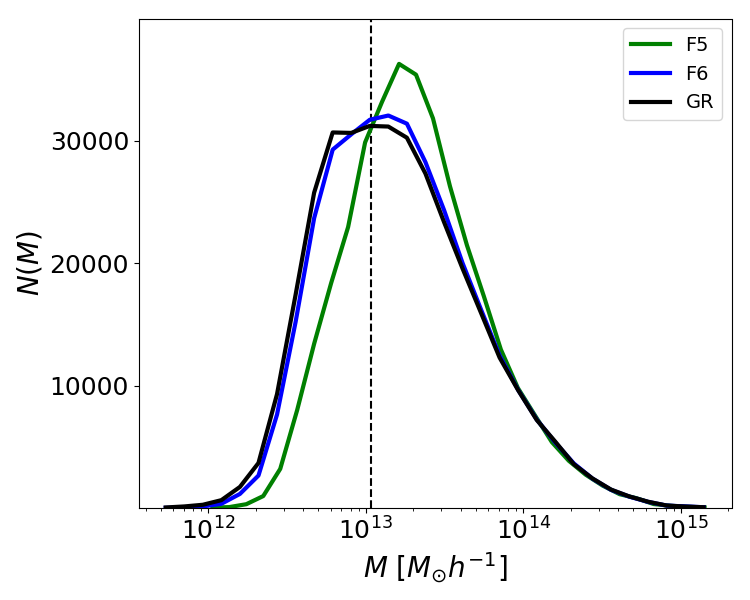}  
\caption{The distribution of the host halo mass $M$ sampled by the HOD galaxies for different models as labelled in the legend. The dashed line indicates mean value for GR.}
\label{fig:m_mark}
\end{figure}

The simulations we employed here were run using the {\sc ecosmog} code \citep{Li:2011vk}, with $1024^3$ dark matter particles with mass $m_p \approx 7.8 \times 10^{10}$ $h^{-1}M_{\odot}$ in a box with size $L = 1024 h^{-1}\text{Mpc}$. We have 5 independent realisations for error analysis. Both $f(R)$ and GR models adopt the same $\Lambda$CDM background cosmology with parameters from the WMAP mission 9-yr results \citep{WMAP9}, hence they essentially have the same expansion history \tcb{and start from identical initial conditions}. Two $f(R)$ models with different amplitude $|f_{R0}|$ are used in this work and are referred to as F5 and F6 (with amplitude values of $|f_{R0}| = 10^{-5},10^{-6}$ respectively). More details can be found in \citet{Cautun2017}. Dark matter haloes were identified by using the {\sc rockstar} code \citep{Behroozi2013} with mass definition $M_{200c}$,\tcb{where the subscript $200c$ refers to 200 times of the critical density of the Universe.}

We populated haloes with galaxies using a 5-parameter halo occupation distribution (HOD) recipe \citep{Zheng:2004id}. The procedure is as follows \citep[see more details in][]{Cautun2017,Li:2017xdi}: 
For GR, we adopted the parameters from \citet{manera2013}, which were calibrated to match the SDSS CMASS clustering. We adjusted the HOD parameters for the $f(R)$ models to best match the galaxy numbers and two-point correlation functions in GR. The flexibility of the HOD model allows us to adjust the shape and magnitude of the galaxy two point correlation function by sampling haloes of different masses, as shown by the histogram for the mass of haloes hosting HOD galaxies by different models in Fig.~\ref{fig:m_mark}. 
This process brought the agreement for the correlation functions among different models to $\leq2\sim3\%$  on scales of between $2-80~h^{-1}$Mpc ({this was calculated as the rms difference between the GR and $f(R)$ correlation functions in all galaxy separation bins, and we also included in the calculation the difference in the galaxy number densities in these models}).

Note the match for the galaxy correlation functions is in real space with no redshift space distortions. This is equivalent to matching the projected two-point correlation functions, as explained in \citet{Cautun2017}. It is also worth noting that the correlation functions agree with each other within the errors estimated {from a volume of $\sim 1(h^{-1}$Gpc)$^3$} of our simulations. 

\begin{figure}
\centering
\includegraphics[width=0.75\linewidth,height=6.5cm]{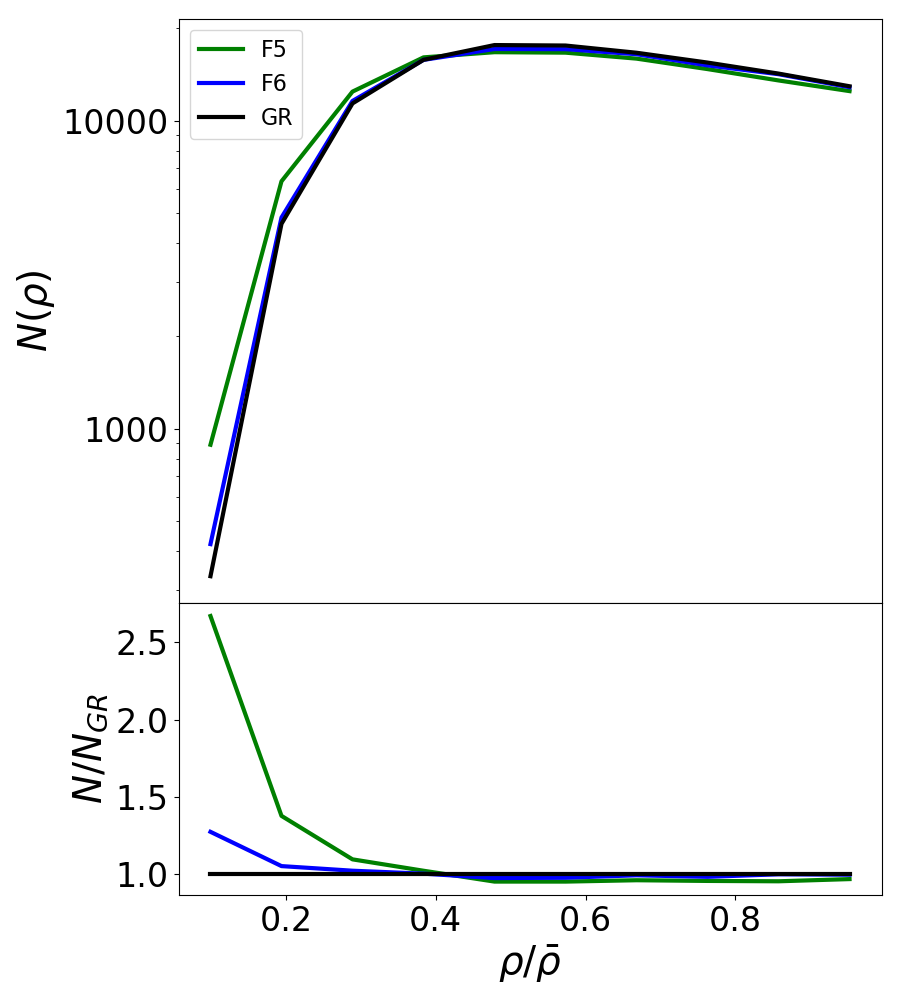} 
\caption{Distribution of galaxy local densities estimated using a Voronoi tessellation method. Only  the range of below the mean density is shown for better illustration.} 
\label{fig:dmark}
\end{figure}

\section{Marked Correlation Function} 
\label{sec:results}

The marked correlation function is in essence a weighted version of the two point correlation function, where the weight is the mark $m$ \citep[e.g.][]{Sheth:2005aj,White2016} 
\begin{equation}
\mathcal{M}(r) = \frac{1}{n(r)\bar{m}^2}\sum_{ij} m_im_j, \label{eq:M_r}
\end{equation}
where $n(r)$ is the number of pairs at separation $r$ in real space, $\bar{m}$ is the mean mark value computed for all the galaxies in the simulation and $m_im_j$ is the product of the marks for the $ij$-galaxy pair. Note that on large scales the average
over all pairs tends toward $\bar{m}^2$, so $\mathcal{M}$ becomes close to unity.

We use the local galaxy number density and the halo mass to define the marks in order to best capture the environmental dependence of structure formation induced by the chameleon screening mechanism in $f(R)$ models.

\subsection{A mark based on local density}

It is well known that for the $f(R)$ model the 5th force is unscreened in low density regions such as voids \citep[e.g.,][]{Hui2009, Clampitt2013}. The consequence is that voids expand faster and become emptier than in GR. The change of large-scale structure in low density regions may not be detectable in the galaxy two point correlation function, which results from the global average of all galaxy pairs. This is because tracers in low density regions have lower amplitudes of clustering by definition, and so their contribution to the total correlation function is minor. As a result, the effect of the chameleon screening may have been hidden under the globally averaged two point correlation function. To amplify the effect due to screening, it is therefore useful to use the local density as a mark, in particular, to up-weight the low density regions.

To do this, we use Voronoi tessellations from the {\sc zobov} code \citep{Neyrinck2008} to estimate the density around each galaxy. The density of a galaxy $\rho_i$ is inversely proportional to the  volume of each Voronoi cell $V_i$. Fig.~\ref{fig:dmark} shows the distribution of galaxy local densities estimated this way. It is clear that while the distributions remain similar to each other for different gravity models for  densities  close to the mean, $f(R)$ models tend to have more galaxies with low densities, i.e. the most isolated galaxies in $f(R)$ models are even more isolated than in GR. In particular, the number of galaxies with $\rho_i <0.2$ could be a factor of 2-3 higher for F5 than for GR. For F6, the difference from GR is milder but the trend is the same. This confirms the expectation that \tcb{the abundance of low density regions is larger in $f(R)$ models} 
even when the galaxy two point correlation functions are the same as in GR. It suggests that having a mark to up-weight the low density regions to enhance this effect may be useful to distinguish $f(R)$ models from GR.

\begin{figure}
\centering
\includegraphics[width=0.8\columnwidth,height=6.8cm]{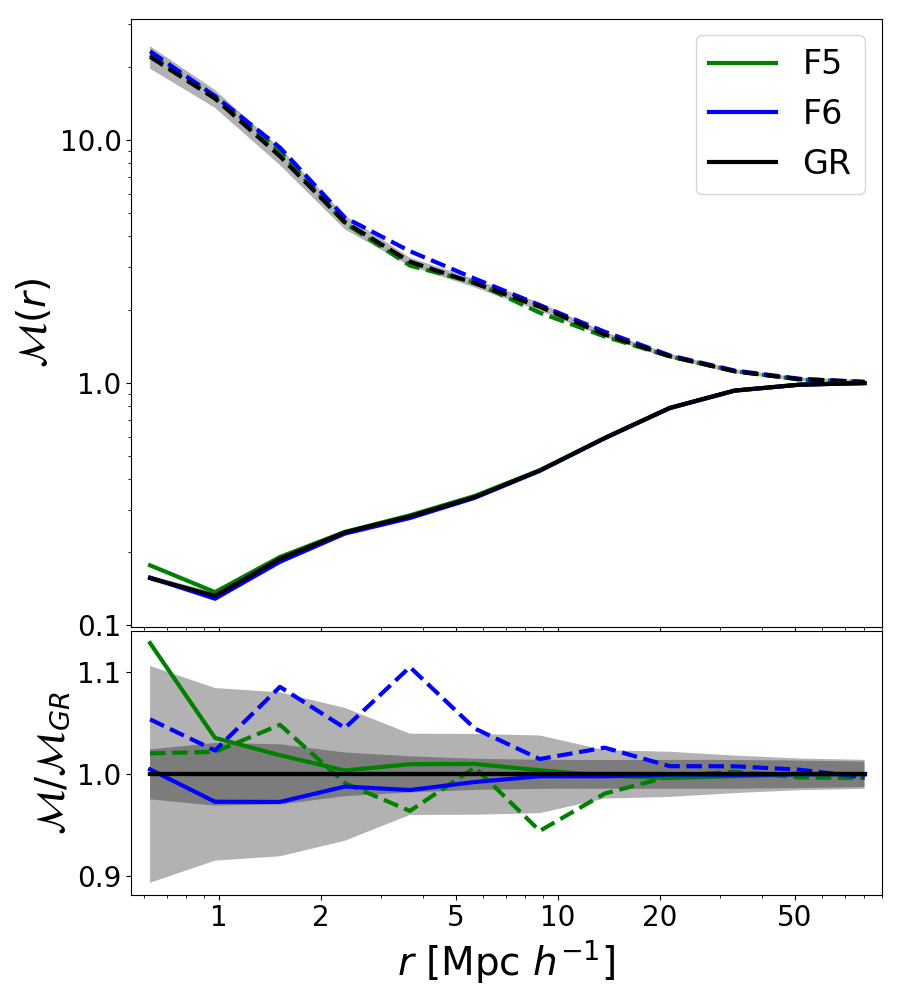}
\caption{The marked correlation function $\mathcal{M}(r)$ using the local density $\rho$ as the mark. This plot shows the examples for $\mathcal M=\rho^{p}$, with $p=\pm 0.5$ in solid (-0.5) and dashed lines (0.5). The lower panel shows the ratios of marked correlation functions between $f(R)$ and GR. The shaded regions correspond to the errors on the mean corresponding to a volume of $\sim 1$($h^{-1}$Gpc)$^3$ estimated using the Jackknife method. The dark and light shaded regions are for the case of $p=-0.5$ and $p=0.5$ respectively.}
\label{fig:d_MCF}
\end{figure}

We first try the mark defined by
$m_i = \rho_i^p$
where the power index $p$ is chosen to be negative to up-weight low density regions. An example for $p=-0.5$ is shown in Fig.~\ref{fig:d_MCF}. For F5, the marked correlation function is above the GR version at the $\sim$2$\sigma$ level at small scales, consistent with the fact that the probability of low density galaxies are higher in this model. For F6 however, it is consistent with GR within the errors, due to the relatively small difference from GR in the distribution function of densities.

These results change with the value of $p$. When $p$ is more negative, e.g. $p<-1$, more weights will be assigned to the low density regions. The relative difference between models 
becomes larger but the noise also increases, because the number of low density galaxies is small. On the other hand, when $p$ is positive, e.g. $p>0.5$, more weights will be assigned to high density regions, which are also rare. In this case, the marked correlations become noisy and indistinguishable from one model to another within the errors. For comparison, an example for $p=0.5$ is also shown in dashed curves in Fig.~\ref{fig:d_MCF}. The light-shaded region in the bottom shows the errors on the mean corresponding to a volume of $\sim$1$(h^{-1}$Gpc)$^3$. \tcb{These errors are estimated using the jackknife method with all the 5 simulation boxes.} The errors are much larger than the case of $p=-0.5$, indicating that the large overdense regions are rarer or higher in their amplitudes than the underdense ones, and so the Poisson noise becomes much larger when up-weighting high densities. Both the F6 and F5 curves are \tcb{broadly consistent with GR} within the errors. \tcb{This confirms the fact that the distribution of galaxies differs more in underdense regions than in overdense regions, and the former carries more information about MG.} \tcb{We have also repeated the same analysis with galaxies in redshift space and find that the marked correlation functions become noisier but results remains qualitatively similar to those in real spcae.}


\subsection{A mark based on halo mass}



\begin{figure*}
 \centering
  \begin{tabular}{c}
   \includegraphics[width=0.95\linewidth]
   {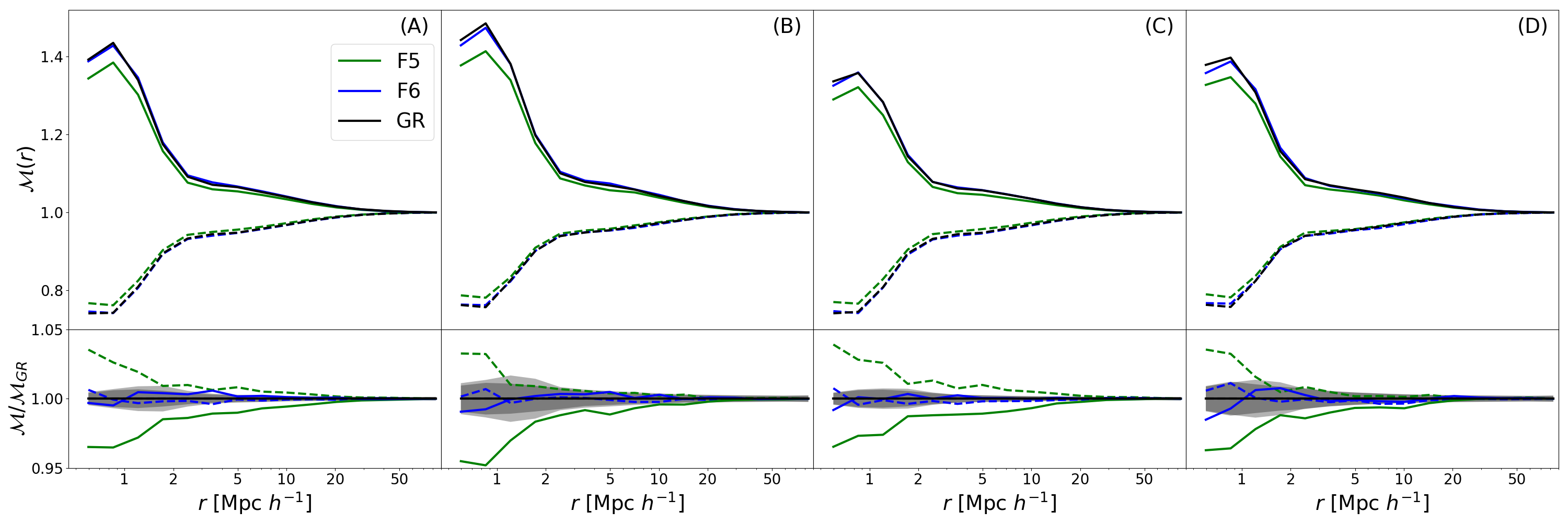} \\
  \end{tabular}
 \caption{Similar to Fig.~\ref{fig:d_MCF} but showing the marked correlation function using the host halo mass of galaxies $M$ as the mark, $\mathcal{M}=M^p$. The solid and dashed curves shows the case for $p=0.1$ and $p=-0.1$ respectively. The dark and light shaded regions show the 1$\sigma$ errors for these two cases. The panels show the different cases: using the host halo mass as mark adding $0.1$~dex uncertainty to the masses (A), adding $0.2$~dex uncertainty (B), using only 8 mass bins to generate the marks and $0.1$~dex uncertainty (C) and adding $0.2$~dex uncertainty (D).} 
\label{fig:m_MCF} 

\end{figure*}

Due to the fifth force, the halo mass functions in $f(R)$ gravity and GR are different \citep[e.g.,][]{Cataneo:2016iav}. The halo occupancies of galaxies therefore have to compensate for this in order to have the same galaxy clustering and number density. This inevitably induces differences in the underlying halo populations being occupied by galaxies, as shown in Fig.~\ref{fig:m_mark}.

Another way to see this is that there are differences in the relations between the galaxy and halo populations in these models. Matching the galaxy density and clustering will result in haloes being populated differently in these models. On the other hand, one can in principle change the HOD parameters such that the halo populations being sampled are the same for different models, but then the galaxy clustering will be different. This difference in the intrinsic relation between haloes and galaxies offers an opportunity to distinguish these two types of models by having a joint constraint from galaxy clustering and their underlying halo population. By using halo mass as the mark in the marked correlation function we can achieve this goal.

To do that, we simply set
$m_i =  M_i^p,$
where $M_i$ is the mass of the host halo, and the index $p$ is a free parameter of our choice. 
We explore a wide range of $p$ and find that F5 can be well distinguished from GR with $0.001<|p|<0.1$. An example for $p=\pm 0.1$ is shown on the left-hand panel in Fig.~\ref{fig:m_MCF}. The marked correlation function for the F5 model deviates from the 1$\sigma$ region of the GR version at scales \tcb{as large as} 20$h^{-1}$Mpc, which is well beyond the 1-halo term region. \tcb{The results remain similar in the above range of $p$}: the amplitude of the marked correlation function decreases with $|p|$, but the errorbars also decrease by approximately the same factor. Therefore, the significance for the deviation from GR is rather independent of $p$. 
When $|p|$ is relatively large, \tcb{i.e. $|p|>0.1$}, the measurement becomes noisy because the tail of the mass distribution is up-weighted \tcb{regardless of the sign of $p$. This is because the distributions of halo mass sampled by the HOD peak at approximately $10^{13}h^{-1}M_{\odot}$ and drops rapidly towards both the low and high mass ends (Fig.~\ref{fig:m_mark})} 
This enhances the Poisson noise and makes F5 indistinguishable from GR \tcb{at $|p|>0.1$}. 
In the limit when $|p|\approx 0$, the mark becomes flat and the correlation functions are equal to unity for all models, and they become indistinguishable from each other. For all the cases we have explored, F6 is always consistent with GR within the errors.

The above experiment suggests a powerful way to constrain the $f(R)$ model, but it requires information about the host halo mass for each galaxy, which is not easily accessible from observation. Even if it is, there will be uncertainties on the halo mass. We therefore make two tests. First, we explore the case where uncertainties for the halo masses are added, i.e. \tcb{$\log_{10}{\tilde{M}_i}=\log_{10}{M_i}+\Delta M$, where $\Delta M=\sigma$ is drawn from a Gaussian distribution with $\sigma$ chosen to be $0.1, 0.2, 0.3$.}
We then measure the marked correlation functions using these noisy marks. \tcb{We find that the results remain qualitatively similar to the case with no noise in terms of the significance for the difference between F5 and GR. As the noise level increases, the errorbars increase as expected. At $\sigma=0.3$, F5 is almost indistinguishable from GR. We show in the panels A \& B of Fig.~\ref{fig:m_MCF} the example for $\sigma=0.1$ \& $0.2$.}

Second, we explore the situation where haloes are binned into 8 mass bins, ranging from $10^{12}$ to $10^{15} h^{-1}M_{\odot}$, with a bin-width of half a decade. \tcb{Note that errors for the halo masses have been added before they are grouped into mass bins.}
The mean mass of host haloes can be estimated either with galaxy-galaxy lensing \citep[e.g.][]{Han2015, Viola2015} or a dynamical method for stacked samples of galaxy groups \citep[e.g.][]{Kaiser1986, Carlberg1997, Evrard2008, Mamon2013}. We then assign galaxies within each mass bin the same mark based on the median mass of the bin, and measure the marked correlation functions. We find that the results remain similar in terms of the differences between the two models, as shown in panels C \& D of Fig.~\ref{fig:m_MCF}. Based on these tests, we conclude that using the halo mass as the mark is a stable and powerful method for distinguishing $f(R)$ and GR models. 

\section{Conclusions and Discussion} 
\label{sec:conc}

We have explored how to use the marked correlation function to distinguish $f(R)$ models from the $\Lambda$CDM universe using N-body simulations. Our study uses different halo occupancies to reproduce the observed projected galaxy two point correlation functions in different models of gravity.
We explore two different marks related respectively to the local galaxy number density and host halo mass, and test their ability to distinguish the models. We find that up-weighting low density regions helps to unveil differences hidden in the correlation function, but only at relatively low significance and on small scales. The latter are actually in the regime of the one-halo term, which can be difficult to interpret in redshift space. Nevertheless, this is qualitatively consistent with the expectation that low-density regions are influenced more strongly by the fifth force in $f(R)$ models. 

\tcb{The method of up-weighting low density regions} is in the same spirit of testing gravity using voids \citep{Clampitt2013,Cai2015}, clipping off peaks \citep{Lombriser2015}, or doing a log transformation on the density \citep{Llinares2017}. \tcb{It also achieves similar goals to the position-dependent power spectrum method in capturing information about three-point statistics \citep{Chiang2014}. } {Our study differs from the recent work of \citet{Valogiannis2017} (VB) where the marked correlation function method was applied to simulations of $f(R)$ and Symmetron models in the following: VB apply the marked statistic to the matter density fields, while we use mock galaxy catalogues, calibrated to have the same clustering and number densities among different models. This sets different requirements for implementing these techniques in observations. 
} 

{We find much stronger deviations between the different models when using halo mass to define the mark. The difference is found out to larger scales ($\sim 20h^{-1}$Mpc) with higher significance. Similar conclusions were found by an independent study  \citep{Cesar:2018} following a similar approach. When using halo mass as the mark we find the result to be stable for a wide range of power indices. The significance remains similar when errors are introduced into the halo mass, or when haloes are grouped into mass bins mimicking stacking to obtain masses via weak lensing, as the method does require additional information about the host halo mass of galaxies. The host halo mass can in principle be measured using a dynamical method or weak gravitational lensing. The latter requires overlapping of a lensing survey and a spectroscopic redshift survey over the same sky. Existing surveys such as GAMA plus KiDS are essentially ready for performing this measurement \citep{Driver2011,Hildebrandt2017}.}

\section*{Acknowledgements}

We thank Peder Norberg and Martin White for helpful discussions. This project has received funding from the European Union's Horizon 2020 Research and Innovation Program under the Marie Sk\l odowska-Curie grant agreement No 734374. JA and NP were supported by proyecto financiamiento BASAL CATA PFB-06 and Fondecyt 1150300. BL is supported by the European Research Council (ERC-StG-716532-PUNCA), and UK STFC Consolidated Grants ST/P000541/1, ST/L00075X/1. This work used the DiRAC Data Centric system at Durham University, operated by the Institute for Computational Cosmology on behalf of the STFC DiRAC HPC Facility (\url{www.dirac.ac.uk}). This equipment was funded by BIS National E-infrastructure capital grant ST/K00042X/1, STFC capital grants ST/H008519/1 and ST/K00087X/1, STFC DiRAC Operations grant ST/K003267/1 and Durham University. DiRAC is part of the National E-Infrastructure. YC and JAP are supported by the European Research Council, under grant no. 670193 (the COSFORM project).

\bibliographystyle{mnras}
\bibliography{references}









\label{lastpage}
\end{document}